\documentclass[12pt,a4paper]{article}
\usepackage[margin=25mm]{geometry}
\usepackage[font=sf]{caption} % Changes font of captions.

\usepackage{amsmath}
\usepackage{amsfonts}
\usepackage{amssymb}
\usepackage{siunitx}
\usepackage{verbatim}

\usepackage[hyphens]{url}
\usepackage[hidelinks]{hyperref}
\usepackage[hyphenbreaks]{breakurl}
\usepackage[T1]{fontenc}
\usepackage{tgtermes}

\usepackage{natbib} % Required for APA-style citations.

\usepackage[dvipsnames]{xcolor}

\title{An international treaty to implement a global compute cap for advanced artificial intelligence}

\author{
  Andrea Miotti\\
  \small{Conjecture}\\
  \small\texttt{andrea@conjecture.dev}}

\date{}

\begin{document}

\maketitle

\begin{abstract}
\noindent
This paper presents an international treaty to reduce risks from the development of advanced artificial intelligence (AI). The main provision of the treaty is a global compute cap: a ban on the development of AI systems above an agreed-upon computational resource threshold. The treaty also proposes the development and testing of emergency response plans, negotiations to establish an international agency to enforce the treaty, the establishment of new communication channels and whistleblower protections, and a commitment to avoid an AI arms race. We hope this treaty serves as a useful template for global leaders as they implement governance regimes to protect civilization from the dangers of advanced artificial intelligence. 

\end{abstract}

\newpage
\tableofcontents

\subsection*{Introduction}

A few artificial intelligence (AI) companies are trying to create extremely powerful AI systems that outperform human reasoning. This kind of technology is sometimes referred to as “artificial general intelligence” (AGI), human-level machine intelligence, artificial superintelligence (ASI), or superhuman machine intelligence.

Many AI experts believe that the development of such AI would have significant threats to public safety, global security, and the continued existence of human civilization. OpenAI CEO Sam Altman has stated that the development of superhuman machine intelligence is “\textbf{the greatest threat to the continued existence of humanity}”, and Anthropic CEO Dario Amodei believes that the chance of an AI-induced civilizational catastrophe is “\textbf{somewhere between 10-25\%}” (Altman, 2015; Oversight of A.I.: Principle for Regulation, 2023a). Amodei has also testified that he believes dangerous AI systems will be created soon: he claimed that systems that enable biological weapons are likely to emerge within \textbf{two to three years} (Oversight of A.I.: Principle for Regulation, 2023a). Yoshua Bengio, recipient of the prestigious Turing Award, states that \textbf{loss of control to rogue AI systems} could occur \textbf{in as little as a few years}, unless appropriate precautions are taken (Oversight of A.I.: Principle for Regulation, 2023b). Finally, many AI experts signed onto a statement earlier this year, which stated that \textbf{risks to human extinction from AI} should be a global priority, much like risks from nuclear war and pandemics (Center for AI Safety, 2023).

Ian Hogarth, the chair of the United Kingdom’s Frontier AI Taskforce, has noted that progress in AI capabilities has substantially outpaced progress in safety research (Hogarth, 2023). He expressed concerns about a “race to God-like AI”: AI companies are aware that the systems they are developing could be highly dangerous, but they are unable to stop racing forward (Hogarth, 2023). Despite the acknowledged chance of catastrophic harm, companies continue developing increasingly more powerful and dangerous AI systems. It seems likely that this dangerous state of affairs will continue unless national or international regulation can halt the race to godlike AI. 

Recently, many AI experts and policymakers have expressed interest in international governance regimes that could end the race to godlike AI systems. Geoffrey Hinton, a recipient of the prestigious Turing Award, has noted that, “We're all in the same boat with respect to the existential threat [from AI], so we all ought to be able to cooperate on trying to stop it” (MIT Technology Review, 2023). Yoshua Bengio, another Turing Award recipient, has called for an international “humanity defense organization” to manage risks from dangerous AI (D’Agostino, 2023). The United Nations (U.N.) Secretary-General Antonio Guterres expressed support for the establishment of an international organization for AI, similar to the International Atomic Energy Agency (IAEA) (Nichols, 2023). 

Experts have emphasized the need for global AI governance measures, international agreements, and international institutions (e.g., Bengio et al., 2023). Scholars have discussed early ideas for what such an international organization could look like and how it might draw from best practices in other fields (Ho et al., 2023; Trager et al., 2023). One concrete and promising proposal involves establishing an international organization that could oversee a moratorium on AI development above a certain amount of computing resources (Hausenloy et al., 2023). Such a proposal could allow for the benefits of safe AI systems to be harnessed and distributed globally while restricting the development of godlike AI systems until experts and policymakers are confident that these can be safely developed and deployed (Hausenloy et al., 2023). 

Despite the widespread interest in international governance regimes, there have been few attempts to draft concrete proposals. Here we present a draft of an international treaty that would implement a global moratorium on dangerous AI development. The treaty incorporates some principles from the Treaty on the Non-Proliferation of Nuclear Weapons (1968), the Chemical Weapons Convention (1993), and a recent proposal for a Treaty on Artificial Intelligence Safety and Cooperation (2023). 

\phantomsection
\addcontentsline{toc}{section}{Treaty Summary}
\subsection*{Treaty Summary}

\textbf{Article 1 (\textit{definitions}) –} defines key terms such as “artificial intelligence”, “compute”, and “advanced hardware.”

\textbf{Article 2 (\textit{general obligations}) –} establishes a Moratorium Threshold; State Parties agree not to develop AI systems beyond the computing power limit set by the Moratorium Threshold. It also establishes a Danger Threshold; states agree to impose regulations on AI developers that develop systems above the danger threshold. Furthermore, states agree to ban the development of human-level AI, AGI, and ASI. 

\textbf{Article 3 (\textit{threshold revisions}) –} acknowledges that the initial criteria for the Moratorium Threshold and the Danger Threshold are imperfect and will need to be lowered over time due to improvements in algorithms and other techniques that make AI systems more powerful.

\textbf{Article 4 (\textit{emergency response plans}) –} State Parties agree to develop and test infrastructure that would allow them to swiftly detect and halt AI development or the proliferation of dangerous AI systems.

\textbf{Article 5 (\textit{monitoring and enforcement}) –} State Parties agree to develop the infrastructure to enforce the obligations set forth by the treaty and to share information about risks with other signatories.

\textbf{Article 6 (\textit{negotiations for creating an international organization for monitoring, enforcement, and research}) –} State Parties agree to engage in good-faith negotiations to create an international organization to enforce the provisions of this treaty, update the Moratorium Threshold and Danger Threshold, and collaborate to develop other measures to reduce the chances of an AI-related global security catastrophe.

\textbf{Article 7 (\textit{sharing the benefits from safe artificial intelligence}) –} State Parties agree to collaborate on establishing measures to ensure that the benefits of safe and beneficial AI are shared globally.

\textbf{Article 8 \textit{(communicating dangers and establishing whistleblower protections}) –} State Parties agree to establish an international hotline to communicate swiftly about potential AI risks or emergencies. State Parties also agree to establish protections for whistleblowers who report unlawful or dangerous AI development practices.

\textbf{Article 9 (\textit{prevention of an artificial intelligence arms race}) –} State Parties agree to engage in good-faith negotiations to prevent an AI arms race.

\textbf{Article 10 (\textit{review conferences}) –} State Parties agree to participate in a review conference at least once a year. The purpose of the conference is to ensure the enforcement of the Treaty is adequate, discuss updates in AI development that may warrant changes or further provisions, and discuss findings from the international agency (described in article 6).

\textbf{Article 11 (\textit{national regulations beyond the scope of the treaty}) –} Clarifies that the Treaty does not prohibit State Parties from implementing their own additional regulations to limit certain kinds of dangerous AI development or deployment.

\textbf{Articles 12 to 14 –} Describe logistics around the implementation of the treaty (e.g., settlement of disputes, signature, ratification, entry into force, withdrawal, and the languages of the authentic texts). 

%--------%
% TREATY %
%--------%

\newpage

\begin{center}

\phantomsection
\addcontentsline{toc}{section}{Treaty on the Prohibition of Dangerous Artificial Intelligence}
\section*{\textbf{TREATY ON THE PROHIBITION OF DANGEROUS ARTIFICIAL INTELLIGENCE }}
\end{center}

\textit{The States Parties to this Treaty},

\textit{Deeply concerned} about the catastrophic consequences that would be visited upon all humankind by a disaster induced by advanced artificial intelligence,

\textit{Acknowledging} the need to make every effort to avert the danger of such a catastrophe and to take measures to safeguard international peace and security,

\textit{Affirming} that artificial intelligence poses risks at least as severe as those from nuclear war, uncontrolled pandemics, and other major threats to global security,

\textit{Believing} that the creation of human-level artificial intelligence or artificial superintelligence should only occur once the international community is confident that such technologies can be controlled and that the necessary national and international governance measures have been established,

\textit{Recognizing} that global security risks from artificial intelligence can occur either from uncontrolled artificial intelligence systems or from human misuse,

\textit{Determined} to eliminate and prevent artificial intelligence race dynamics between countries and between corporations which significantly raise the risk of catastrophe,

\textit{Acknowledging} the benefits that advanced artificial intelligence could bring to humanity once there is greater certainty that such technology can be developed and governed safely,

\textit{Expressing} their support for research, development, and other efforts to safeguard the production and trade of powerful artificial intelligence hardware and to identify privacy-preserving methods of monitoring compliance with hardware regulations, 

\textit{Reaffirming} the United Nation’s commitment to achieve international co-operation in solving international problems of an economic, social, cultural, or humanitarian character,

\textit{Urging} the cooperation of all States in the prevention of catastrophes caused by artificial intelligence,

\textit{Desiring} to further facilitate the monitoring of advanced hardware, the avoidance of an artificial intelligence arms race, and the elimination and prevention of efforts to prematurely develop human-level artificial intelligence, artificial superintelligence, and other forms of highly dangerous artificial intelligence, 

\textit{Have agreed} as follows:

\begin{center}
    
\textbf{{ARTICLE I}}

\textit{\textbf{Definitions}}
\end{center}

For the purposes of this Treaty:

1. “Advanced hardware” means powerful computing semiconductor chips or integrated circuits that can be used to build artificial intelligence systems above the Danger Threshold. 

2. “Algorithmic improvement” means advancements in artificial intelligence algorithms, methodologies, architectures, or techniques that lead to either: (a) a reduction in the computational resources, data, time, or cost required to develop advanced artificial intelligence systems or (b) an improvement in artificial intelligence capabilities.

3. “Artificial intelligence” means the following, together or separately:

\begin{enumerate}
    \item[(a)] Any artificial system that performs tasks under varying and unpredictable circumstances without significant human  oversight, or that can learn from experience and improve performance when exposed to data sets.
    \item[(b)] An artificial system developed in computer software, physical hardware, or other context that solves tasks requiring human-like perception, cognition, planning, learning, communication, or physical action.
    \item[(c)] An artificial system designed to think or act like a human, including cognitive architectures and neural networks.
    \item[(d)]A set of techniques, including machine learning, that is designed to approximate a cognitive task.
    \item[(e)]An artificial system designed to act rationally, including an intelligent software agent or embodied robot that  achieves goals using perception, planning, reasoning, learning, communicating, decision-making, and acting.
    \item[(f)] A machine-based system that is capable of influencing the environment by producing an output (predictions, recommendations or decisions) for a given set of objectives. It uses machine and/or human-based data and inputs to (i) perceive real and/or virtual environments; (ii) abstract these perceptions into models through analysis in an automated manner (e.g., with machine learning), or manually; and (iii) use model inference to formulate options for outcomes. 
\end{enumerate}

4. “Artificial general intelligence” or “human-level artificial intelligence” means artificial intelligence that achieves human-level performance at a wide variety of intellectual tasks, without being constrained to a narrow or specific domain of expertise.

5. “Artificial superintelligence” means artificial intelligence that exceeds human-level performance in most or all domains, potentially including general problem-solving, social skills, planning and strategic thinking, scientific research, and artificial intelligence development. 

6. “Compute” means the processing power and other electronic resources used to train, validate, deploy, and run artificial intelligence algorithms and models. 

7. “Dangerous artificial intelligence systems” includes collectively “human-level artificial intelligence”, “artificial general intelligence”, “artificial superintelligence”, and any artificial intelligence systems that pose severe global or national security risks.

8. “Floating-Point Operations” (FLOP) means single-precision (32-bit) floating point operations.

9. “Whistleblower” means any individual who reports unlawful or dangerous artificial intelligence development practices to the State within whose borders the unlawful or dangerous artificial development practices are carried out or to a trusted international agency

 \begin{center}
    
\textbf{{ARTICLE II}}

\textit{\textbf{General Obligations}}
\end{center}

1. Each State Party undertakes to prohibit the civilian or military development, deployment, transfer, possession and use of artificial intelligence systems above the Moratorium Threshold. 

2. Each State Party undertakes to regulate the development and use of artificial intelligence systems above the Danger Threshold but below the Moratorium Threshold such that any entity developing or using systems above the Danger Threshold (but below the Moratorium Threshold) must show that they are following appropriate regulations and safeguards. Examples include information security requirements, probabilistic risk assessments, predictions of dangerous capabilities, third-party auditing, and other regulations protecting safety and fundamental rights.

3. The Moratorium Threshold and the Danger Threshold shall initially be set based on the compute required to develop artificial intelligence systems (which serve as a proxy for the model's capabilities). The Moratorium Threshold will start at 10\^{}24 Floating-Point Operations and the Danger Threshold will start at 10\^{}21 FLOP. 

4. Each State Party undertakes to prohibit the development and use, for civilian or military purposes, of human-level artificial intelligence, artificial general intelligence (AGI), artificial superintelligence (ASI), or other forms of highly dangerous artificial intelligence systems. 

5. Each State Party undertakes to implement comprehensive regulations to monitor artificial intelligence development, to report any instances in which an individual or group is suspected of developing or using one or many highly dangerous artificial systems or is suspected to be attempting to develop or use one or many highly dangerous artificial intelligence systems to the United Nations Security Council, and to investigate credible threats.

\begin{center}
    
\textbf{{ARTICLE III}}

\textit{\textbf{Threshold Revisions}}
\end{center}

1. One year after the entry into force of this Treaty, a conference of Parties shall be held in Geneva, Switzerland, in order to review the Moratorium Threshold.  

2. At an interval at least once per year the State Parties will meet in Geneva, Switzerland, in order to review the Moratorium Threshold. The definitions of the Moratorium Threshold and Danger Threshold are recognized to be imperfect proxies that will be updated over time. It is further recognized that the development, acquisition, possession, or use of dangerous artificial intelligence systems will require fewer computing resources over time, due to algorithmic progress and other improvements (e.g., the discovery of new prompting techniques) to artificial intelligence development. 

 \begin{center}
    
\textbf{{ARTICLE IV}}

\textit{\textbf{Emergency response plans}}

\end{center}

1. Each State Party commits to developing and testing one or more emergency response plans in which States demonstrate the capacity to swiftly detect and halt dangerous artificial intelligence development (e.g., stop a training run before or immediately after it crosses the Moratorium Threshold) or stop the proliferation of a dangerous artificial intelligence model (e.g., withdraw application programming interface (API) access). 

2. Each State Party undertakes to conduct tests of its emergency response plan or plans at regular intervals to ensure that States have the capacity to respond effectively in the event of an emergency. 

3. Each State Party agrees to share information in good faith relating to the monitoring of artificial intelligence capabilities and development. 

  \begin{center}
    
\textbf{{ARTICLE V}}

\textbf{\textit{Monitoring and enforcement}}

\end{center}

1. Each State Party undertakes to take appropriate measures to ensure the enforcement of this treaty, including the development of the infrastructure required to enforce the Treaty.

2. Each State Party undertakes to self-report the amount and locations of large concentrations of advanced hardware to relevant international authorities. 

3. Each State Party recognizes that the self-reporting procedure must allow for the comprehensive verification of advanced hardware in declared facilities in order to monitor that it is not being used to develop artificial intelligence above the Moratorium Threshold and to allow for the detection of any undeclared or secret facilities with large concentrations of advanced hardware.

4. Each State Party recognizes the need for the establishment of a protocol allowing for investigation by independent evaluators within their borders and undertakes to negotiate in good faith to that effect. 

\begin{center}
    
\textbf{{ARTICLE VI}}

\textbf{\textit{Negotiations for creating an international organization for monitoring, enforcement, and research }}
\end{center}

1. Each State Party undertakes to engage in good-faith negotiations to create an international agency for the purpose of verification of the fulfillment of its obligations assumed under this Treaty. The primary purpose of this international agency would be to ensure that the provisions in the Treaty are effectively enforced. 

2. The agency would also be responsible for researching highly powerful artificial intelligence systems, with the ultimate goal of understanding how to control highly powerful artificial intelligence systems and ensure that they are only ever developed for the benefit of the whole of humanity. 

3. The agency would be responsible for adjusting the Moratorium Threshold and the Danger Threshold. The Moratorium Threshold may be lifted if the agency acquires compelling evidence that they are able to safely build and deploy human-level artificial intelligence and artificial superintelligence. 

 \begin{center}
    
\textbf{{ARTICLE VII}}

\textbf{\textit{\textit{\textbf{\textit{\textbf{Sharing the benefits from safe artificial intelligence}} }}}}
\end{center}

Each State Party undertakes to collaborate in good-faith for the establishment of effective measures to ensure that potential benefits from safe and beneficial artificial intelligence systems are distributed globally. 

\begin{center}
    
\textbf{{ARTICLE VIII}}

\textit{\textbf{Communicating dangers and establishing whistleblower protections}} 
\end{center}
1. Each State Party undertakes to establish and participate in an international hotline, allowing for direct communication between leaders in each State pertaining to matters of global security threats related to artificial intelligence. 

2. Each State Party agrees to report evidence of dangerous artificial intelligence development, insights about dangerous capabilities, suspected noncompliance with the Treaty, and any other issue posing a threat to global security. 
3. Each State Party undertakes to establish a similar communication channel for civilian artificial intelligence developers. Civilian artificial intelligence developers will be required to share evidence of dangerous artificial intelligence development, insights about dangerous capabilities, suspected noncompliance with the Treaty, and any information on any issue posing a potential threat to global security.

3. Each State Party undertakes to establish appropriate protections for whistleblowers who report unlawful or dangerous artificial intelligence development practices to the State or a trusted international entity. 

 \begin{center}
    
\textbf{{ARTICLE IX}}

\textbf{\textit{Prevention of an artificial intelligence arms race}}
\end{center}

Each State Party undertakes to pursue in good faith negotiations on effective measures relating to the cessation of an artificial intelligence arms race and the prevention of any future artificial intelligence arms race. 
    
 \begin{center}
    
\textbf{{ARTICLE X}}

\textbf{\textit{Review conferences}}
\end{center}

1. At least once per year after the entry into force of this Treaty, a conference of the State Parties shall be held in Geneva, Switzerland. 

2. The purpose of the conference will be to review the operation of this Treaty to assure that the purposes of the Preamble and the provisions to the Treaty are being realized, to discuss possible changes to the Moratorium Threshold and Danger Threshold to account for artificial intelligence progress (described in Article II), and discuss other matters pertaining to global security threats from artificial intelligence. 

\begin{center}
    
\textbf{{ARTICLE XI}}

\textbf{\textit{National regulations beyond the scope of the treaty}}
\end{center}

1. Nothing in this Treaty affects the right of any State or group of States to implement regulations based on definitions that include criteria other than a FLIP threshold (such as benchmark performance, parameter count, the domains in which the artificial intelligence can be applied, or the presence of certain dangerous capabilities). 

2. Nothing in this Treaty affects the right of any State or group of States to implement regulations on artificial intelligence systems below the Moratorium and Danger Thresholds, or additional regulations applied to systems above the Danger Threshold (but below the Moratorium Threshold) that do not interfere with the obligations provided for by the Treaty.

3. Each State Party accepts the responsibility to impose its own regulations on artificial intelligence developers who are engaging, under said State Party’s jurisdiction, in activities that could pose national or global security threats.

\begin{center}
    
\textbf{{ARTICLE XII}}

\textbf{\textit{Settlement of disputes}}
\end{center}

When a dispute arises between two or more States Parties relating to the interpretation or application of this Treaty, the parties concerned shall consult together with a view to the settlement of the dispute by negotiation or by other peaceful means of the parties’ choice in accordance with Article 33 of the Charter of the United Nations. 

\begin{center}
    
\textbf{{ARTICLE XIII}}

\textbf{\textit{Signature, ratification, entry into force, and withdrawal}}
\end{center}

1. This Treaty shall be open for signature to all States for signature before its entry into force. 

2. This Treaty shall be subject to ratification, acceptance or approval by signatory States. The Treaty shall be open for accession.

3. This Treaty shall enter into force 60 days after the date of the deposit of the second instrument of ratification. 

4. For States whose instruments of ratification, acceptance, approval or accession are deposited subsequent to the entry into force of this Treaty, it shall enter into force on the 30th day following the date of deposit of their instrument of ratification, acceptance, approval or accession.

5. Each State Party shall, in exercising its national sovereignty, have the right to withdraw from this Treaty if it decides that extraordinary events related to the subject matter of the Treaty have jeopardized the supreme interests of its country. It shall give notice of such withdrawal to the Depositary. Such notice shall include a statement of the extraordinary events that it regards as having jeopardized its supreme interests. Such withdrawal will only take effect 6 months after the date of receipt of the notification of withdrawal. 

6. The Secretary-General of the United Nations is hereby designated as the Depositary of this Treaty.

\begin{center}
    
\textbf{{ARTICLE XIV}}

\textbf{\textit{Authentic texts}}
\end{center}
The Arabic, Chinese, English, French, Russian and Spanish texts of this Treaty shall be equally authentic.

IN WITNESS THEREOF the undersigned, being duly authorised, have signed this Treaty.

\newpage
\bibliographystyle{apalike}
\bibliography{example}
\sloppy

Altman, S. (2015). \textit{Machine Intelligence, part 1}. \url{https://blog.samaltman.com/machine-intelligence-part-1}

Bengio, Y., Hinton, G., Yao, A., Song, D., Abbeel, P., Harari, Y. N., Zhang, Y.-Q., Xue, L., Shalev-Shwartz, S., Hadfield, G., Clune, J., Maharaj, T., Hutter, F., Baydin, A. G., McIlraith, S., Gao, Q., Acharya, A., Krueger, D., Dragan, A., Torr, P., Russell, S., Kahneman, D., Brauner, J., \& Mindermann, S. (2023). \textit{Managing AI Risks in an Era of Rapid Progress}. \url{https://arxiv.org/abs/2310.17688}

Center for AI Safety (CAIS) (2023). \textit{Statement on AI Risk}. \url{https://www.safe.ai/statement-on-ai-risk}

\textit{Chemical Weapons Convention}, January 13, 1993. Organisation for the Prohibition of Chemical Weapons. \url{https://www.opcw.org/chemical-weapons-convention}

D’Agostino, S. (2023). \textit{‘AI Godfather’ Yoshua Bengio: We need a humanity defense organization}. \url{https://thebulletin.org/2023/10/ai-godfather-yoshua-bengio-we-need-a-humanity-defense-organization/}

Hausenloy, J., Miotti, A. \& Dennis, C. (2023). \textit{Multinational AGI Consortium (MAGIC): A Proposal for International Coordination on AI}. \url{https://arxiv.org/abs/2310.09217}

Ho, L., Barnhart, J., Trager, R., Bengio, Y., Brundage, M., Carnegie, A., Chowdhury, R., Dafoe, A., Hadfield, G., Levi, M. \& Snidal, D. (2023). \textit{International Institutions for Advanced AI}. \url{https://arxiv.org/abs/2307.04699}

Hogarth, I. (2023) \textit{We must slow down the race to God-like AI}. \url{https://www.ft.com/content/03895dc4-a3b7-481e-95cc-336a524f2ac2}

MIT Technology Review. (2023).\textit{ Video: Geoffrey Hinton talks about the “existential threat” of AI.} \url{https://www.technologyreview.com/2023/05/03/1072589/video-geoffrey-hinton-google-ai-risk-ethics/}

Nichols, M. (2023). UN chief backs idea of global AI watchdog like nuclear agency. \textit{Reuters}. \url{https://www.reuters.com/technology/un-chief-backs-idea-global-ai-watchdog-like-nuclear-agency-2023-06-12/}

\textit{Oversight of A.I.: Principles for Regulation: Hearing before the Judiciary Committee Subcommittee on Privacy, Technology, and the Law}, U.S. Senate, 118th Congr. (2023a). (testimony of Dario Amodei). \url{https://www.judiciary.senate.gov/imo/media/doc/2023-07-26_-_testimony_-_amodei.pdf}

\textit{Oversight of A.I.: Principles for Regulation: Hearing before the Judiciary Committee Subcommittee on Privacy, Technology, and the Law}, U.S. Senate, 118th Congr. (2023b). (testimony of Yoshua Bengio). \url{https://www.judiciary.senate.gov/imo/media/doc/2023-07-26_-_testimony_-_bengio.pdf}

Trager, R., Harack, B., Reuel, A., Carnegie, A., Heim, L., Ho, L., Kreps, S., Lall, R., Larter, O., Ó hÉigeartaigh, S., Staffell, S. \& Villalobos, J. (2023). \textit{International Governance of Civilian AI: A Jurisdictional Certification Approach}. \url{https://papers.ssrn.com/sol3/papers.cfm?abstract_id=4579899}

\textit{Treaty on Artificial Intelligence Safety and Cooperation} [Treaty Blueprint]. (2023). \url{https://taisc.org/taisc}

\textit{Treaty on the Non-Proliferation of Nuclear Weapons}, 1 July 1968. \url{https://treaties.unoda.org/t/npt}

\end{document}